\documentclass[3p,times,procedia]{elsarticle}
\usepackage{nupha_ecrc}

\volume{00}

\firstpage{1}

\journalname{Nuclear Physics A}

\runauth{}

\jid{nupha}

\jnltitlelogo{Nuclear Physics A}

\usepackage{amssymb}

\usepackage[figuresright]{rotating}

\begin{document}

\begin{frontmatter}



\dochead{XXVIIth International Conference on Ultrarelativistic Nucleus-Nucleus Collisions\\ (Quark Matter 2018)}

\title{Transport properties from Charm to Bottom: $p_T$ suppression, anisotropic flow $v_n$ and their correlations to the bulk dynamics}


\author[L1]{S. Plumari}
\author[L1,L2]{G. Coci}
\author[L3,L1]{S.K. Das}
\author[L1,L2]{V. Minissale}
\author[L1,L2]{V. Greco}

\address[L1]{Department of Physics and Astronomy, University of Catania, Via S. Sofia 64, 1-95125 Catania, Italy}
\address[L2]{Laboratori Nazionali del Sud, INFN-LNS, Via S. Sofia 62, I-95123 Catania, Italy}
\address[L3]{School of Nuclear Science and Technology, Lanzhou University, 730000 Lanzhou, China}

\begin{abstract}
  We study the propagation of heavy quarks (HQs) in the quark-gluon plasma (QGP) by means of a relativistic Boltzmann transport (RBT) approach. The non-perturbative interaction between heavy quarks and light quarks is described by means of a quasi-particle approach able to describe simultaneously the experimental data for the nuclear suppression factor $R_{\rm AA}$ and the elliptic flow $v_2(p_T)$ of D mesons from RHIC to LHC energies. In the same framework we predict the B meson nuclear modification factor at LHC.
  Finally, we discuss the relevance of initial state fluctuations that allows to extend the analysis to high order anisotropic flows $v_n(p_T)$ as well as to investigate the role of QCD interaction in developing correlations between the light and the heavy flavour anisotropic flows.
\end{abstract}


\end{frontmatter}


\section{Introduction}
\label{section0}
Heavy quarks (HQs), charm and bottom quarks, represent excellent probes
of the system created at ultra-Relativistic Heavy Ion Collisions (uRHIC).
Their formation time is very small compared to the light quarks one,
and due to their large masses they are expected to thermalize slower in
the Quark-Gluon Plasma (QGP). Therefore, HQ can probe both for the initial
stages of uRHIC and the thermalized QGP evolution. In their final state the
charm quarks appear as constituent of charmed and bottom hadrons mainly $D$, $B$ mesons and
$\Lambda_{c}$, $\Lambda_{b}$ baryons. 
Two key observables in HQ sector are the nuclear suppression factor
$R_{AA}$ (the ratio between the spectra of heavy flavour hadrons in nucleus-nucleus 
collisions and the one in proton-proton collisions) and the elliptic flow $v_2(p_T)$
(a measure of the anisotropy in momentum space).
Experimental measurements, both at RHIC and LHC, have shown many interesting observations
for heavy mesons $R_{AA}$ and $v_2(p_T)$. In particular it was observed a small $R_{AA}$ value
and large value of $v_2(p_T)$, which are almost comparable to those of light hadrons.
Several theoretical efforts have been made to study the $R_{AA}$ and the $v_2$ within different models
~\cite{vanHees:2007me,Alberico:2011zy,Cao:2015hia,Uphoff:2011ad,Berrehrah:2014kba,Song:2015ykw,Das:2013kea,Das:2015ana,Beraudo:2014boa,Scardina:2017ipo}.
Besides the well studied $v_2(p_T)$, it has been recently shown that also the triangular flow $v_3(p_T)$ 
of D mesons is not vanishing \cite{Nahrgang:2014vza,Beraudo:2017gxw}.
In recent years, it has been recognized that a strong Electromagnetic (EM) field is created at early
times of uRHIC. Since HQs are produced in the very early stages of uRHICs they will be directly affected by such
a strong EM field and this results in a rapidity-odd directed flow $v_1$ for $D^0$ and $\bar{D^0}$ \cite{Das:2016cwd}.
\section{Transport equation for charm quarks in the QGP}
\label{section1}
We describe the charm quarks evolution in the QGP
by solving the RBT equations where 
charm quarks interacts with a bulk medium of quarks 
and gluons as described by the following eq.s 
\begin{eqnarray}
 p^{\mu} \partial_{\mu}f_{Q}(x,p)&=& {\cal C}[f_q,f_g,f_{Q}](x,p)  \label{Eq:BE1} \\
 p^{\mu} \partial_{\mu}f_{j}(x,p)&=& {\cal C}[f_q,f_g](x_q,p_q)  \, \hspace{2pc} \, j=q,g \label{Eq:BE2} 
\end{eqnarray}
where $f_j(x,p)$ is the phase-space one-body distribution function of the $j$ parton (quark, anti-quark or gluon) 
while ${\cal{C}}[f_q, f_g, f_{Q}](x,p)$ refers to the relativistic Boltzmann-like collision integral. 
As shown in Eq.(\ref{Eq:BE1}) the phase-space distribution function of the bulk medium (quarks and gluons) enters in 
the evolution equation for charm quarks as an external quantities with ${\cal{C}}[f_q,f_g,f_{Q}]$, and  
the evolution of $f_q$ and $f_g$ have been assumed to be independent of $f_{Q}(x,p)$.
We discard collisions between heavy quarks which is by far a solid approximation.
The evolution of the bulk is given by the two equations Eqs.(\ref{Eq:BE2}), 
where in ${\cal C}[f_q,f_g]$ the total cross section is determined in order to keep fixed the ratio $\eta/s=1/(4\pi)$
during the evolution of the QGP, for a detailed discussion 
see ref.s~\cite{Plumari:2012xz,Ruggieri:2013ova,Plumari:2015cfa}.
The non-perturbative interaction between heavy quarks and light quarks is described by means of a quasi-particle
approach \cite{Plumari:2011mk}. This provides a softening of the equation of state, with a decreasing speed of sound
approaching the cross-over region.
In this approach we describe the evolution of a system that dynamically has approximatively the lQCD equation of state
\cite{Borsanyi:2010cj}.
The hadronization process plays a crucial role in determining the final spectra, $R_{\rm AA}(p_T)$ and $v_2(p_T)$.
When the temperature of the QGP phase goes below the quark-hadron transition temperature, $T_c=155$ MeV, we hadronize 
the charm quark to D-meson. We have considered a hybrid model of coalescence plus 
fragmentation (for a detailed discussion of the hadronization model see \cite{Plumari:2017ntm}).
We have studied $Au+Au$ collisions at $\sqrt{s_{\rm NN}}= 200 $ GeV at RHIC and
$Pb+Pb$ collisions at $\sqrt{s_{\rm NN}}= 2.76 $ TeV at LHC.
The initial conditions for the bulk in the coordinate space are given by the standard Glauber model
assuming boost invariance along the longitudinal direction.
In momentum space are given by a Boltzmann-Juttner distribution function up to a transverse momentum $p_T=2$ GeV while
at larger momenta mini-jet distributions as calculated by pQCD at NLO order in \cite{Greco:2003xt}. 
The initial temperature at the center of the fireball is fixed to $T_0=365 $ MeV with the initial 
time for the simulations $\tau_0 \simeq 1/T_0= 0.6 \,fm/c$ for RHIC and 
$T_0=490 $ MeV with $\tau_0 \simeq 1/T_0= 0.3 \,fm/c$ at LHC.
In coordinate space we initialize the charm quark distribution according 
to the number of binary nucleon-nucleon collisions, $N_{coll}$. 
In momentum space we use charm quark production according to the Fixed Order + Next - to - Leading Log (FONLL)
calculation (from Ref.~\cite{Cacciari:2012ny}) which describes the D-mesons spectra in proton-proton collisions after fragmentation Ref.~\cite{Scardina:2017ipo}. 
\section{Results}
\label{Intro}
\begin{figure}
\begin{center}
  \includegraphics[width=.36\textwidth]{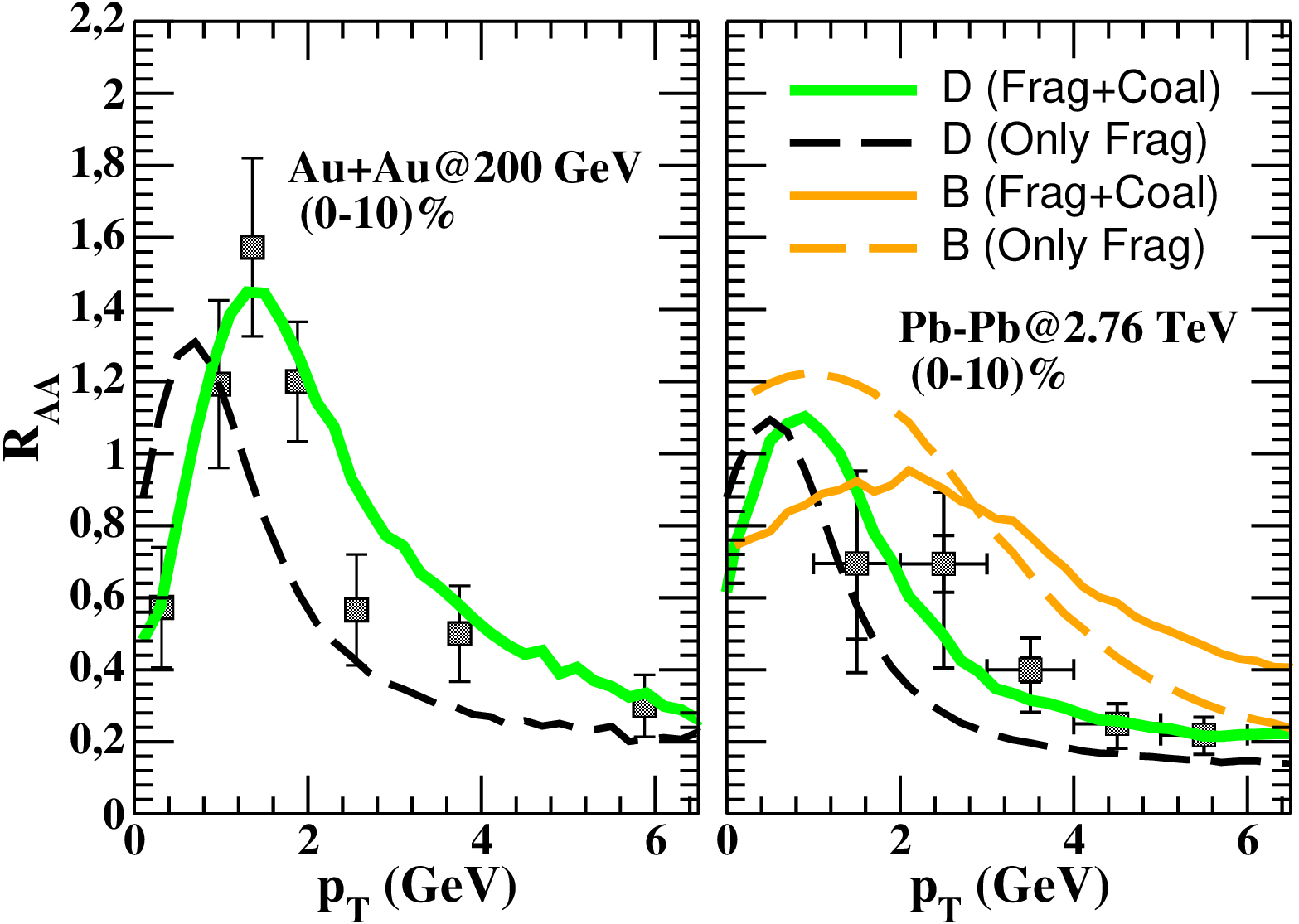}
\hspace{1.5pc}
\includegraphics[width=.35\textwidth]{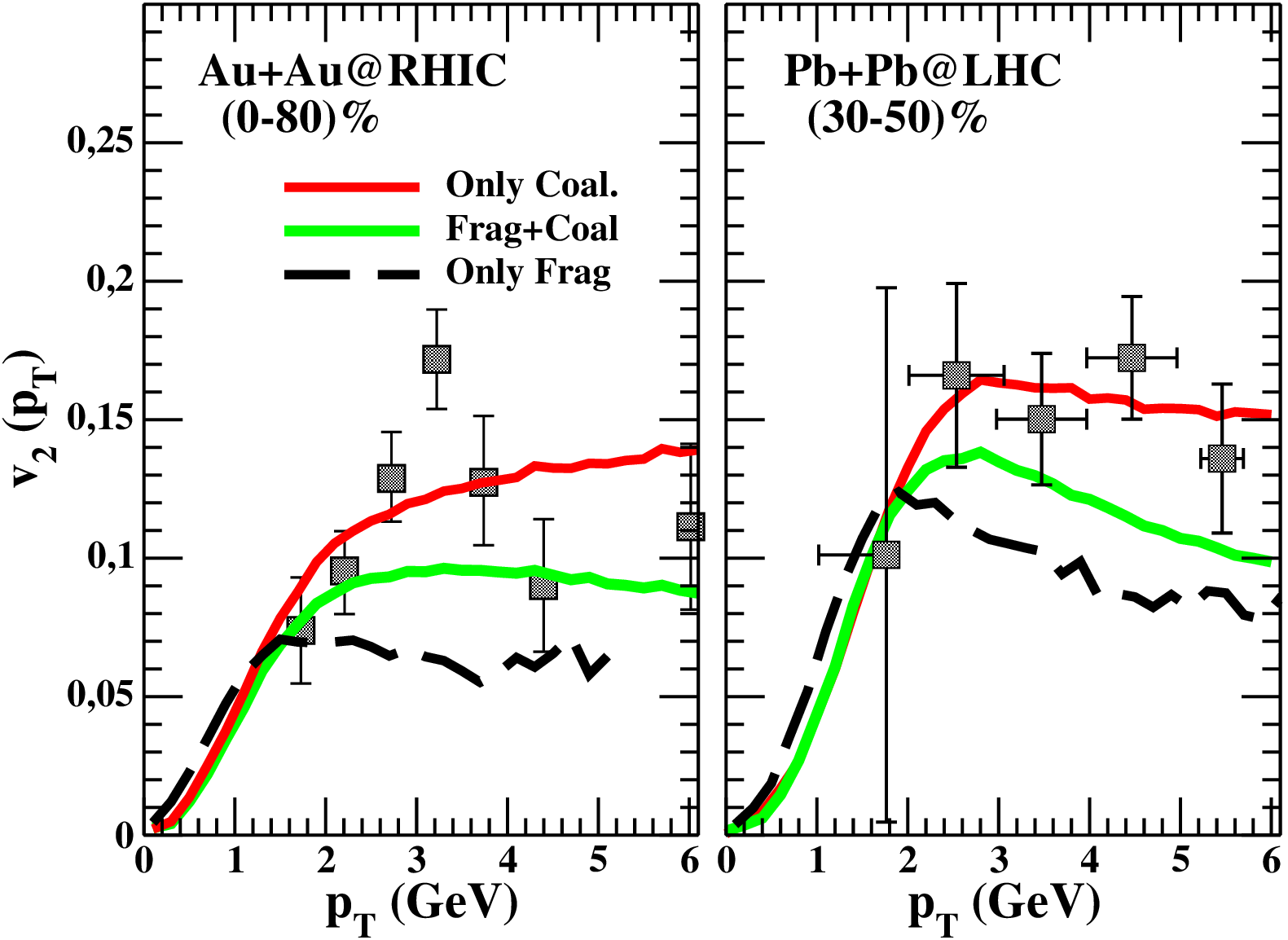}
\caption{
  Left panel: D meson $R_{\rm AA}$ for ($0-10 \%$) at RHIC in $Au+Au$ collisions at $\sqrt{s_{\rm NN}}=200 $ GeV and LHC in $Pb+Pb$ collisions at $\sqrt{s_{\rm NN}}=2.76 $ TeV compared to the experimental data (Ref.s ~\cite{Adamczyk:2014uip,Adam:2015sza}). The orange dashed and solid lines refer to the B meson $R_{\rm AA}$ with only fragmentation and with both coalescence plus fragmentation respectively. Right panel: $v_2$ at RHIC energy for ($0-80\%$) and at LHC energy for ($30-50\%$) compared to the experimental data (Ref.s ~\cite{Adamczyk:2017xur,Abelev:2013lca}).The black dashed lines and red lines refer to the case with only fragmentation and only coalescence respectively. The green solid lines to the case with fragmentation plus coalescence.
}\label{fig1}
\end{center}
\end{figure}
We have calculated the $R_{\rm AA}$ and $v_2(p_T)$ of D meson at different centrality class from RHIC to LHC energies, using the same interaction.
Our model gives a good description for D mesons $R_{\rm AA}$ and $v_2(p_T)$ at both energies. 
In the left panel of Fig.~\ref{fig1} it is shown the comparison of our results for the $R_{\rm AA}(p_T)$ with the experimental data 
for both systems created at RHIC and LHC energies for ($30-50\%$) centrality.
We observe by comparing black dashed lines with green solid lines, that the effect of coalescence is to increase the $R_{\rm AA}$ 
for momenta larger than $1 $ GeV.
For the hadronization mechanism by coalescence, D mesons, which are composed by one light quark and a charm quark, get a larger momentum with respect to 
the D mesons obtained from fragmentation.
On the other hand, at larger momenta, the coalescence contribution to hadron formation becomes very small and fragmentation becomes,
anyway, the dominant mechanism of hadronization.
From the comparison at different energies, the effect of coalescence
is less significant at LHC than at RHIC.
This is because the effect of coalescence depends on the slope of HQ spectrum. For an harder charm quark distribution, like at LHC energy, the impact of coalescence is therefore less pronounced (for details see \cite{Scardina:2017ipo,Plumari:2017ntm}).
We note that, if $\Lambda_c$ is included the D meson $R_{\rm AA}$ will be substantially modified at low $p_T$.
We have also shown in Fig.~\ref{fig1}, by the orange dashed and solid lines, the B meson $R_{\rm AA}$ with only fragmentation and with
both fragmentation plus coalescence. Due to their larger masses, bottoms quarks lose a smaller amount of energy than charm quark
and therefore the B meson displays larger $R_{\rm AA}$ than D mesons. Moreover, due to the fact that for bottom quarks is easier to combine with light quarks, the effect of coalescence in B meson $R_{\rm AA}$ is more evident than for D meson $R_{\rm AA}$.
In the right panel of Fig.~\ref{fig1}, we show the corresponding results for the final $v_2(p_T)$ of D mesons
for both RHIC and LHC in mid-peripheral collisions.
As shown, the $v_2$ developed via only coalescence (red solid lines) is larger than the $v_2$ due to fragmentation (black dashed lines).
This difference cames from the fact that the D meson elliptic flow formed via coalescence reflects both the heavy quark and light quark anisotropies 
in momentum space and it can even lead to an increase of about a factor two at $p_T>2 $ GeV.
Finally, as shown by green solid lines, when one consider the hybrid hadronization by coalescence plus fragmentation,
the $v_2$ of the D-mesons increases with respect to the $v_2$ of D meson from only fragmentation by about a $30\%$.
The coalescence play a key role to get a good description of the experimental data and moreover the extracted diffusion
coefficient $2\pi T D_s$ is in agreement with lattice QCD results at least within the current uncertainties (for a detailed discussion see \cite{Scardina:2017ipo}).
\begin{figure}
\begin{center}
  \includegraphics[width=.35\textwidth]{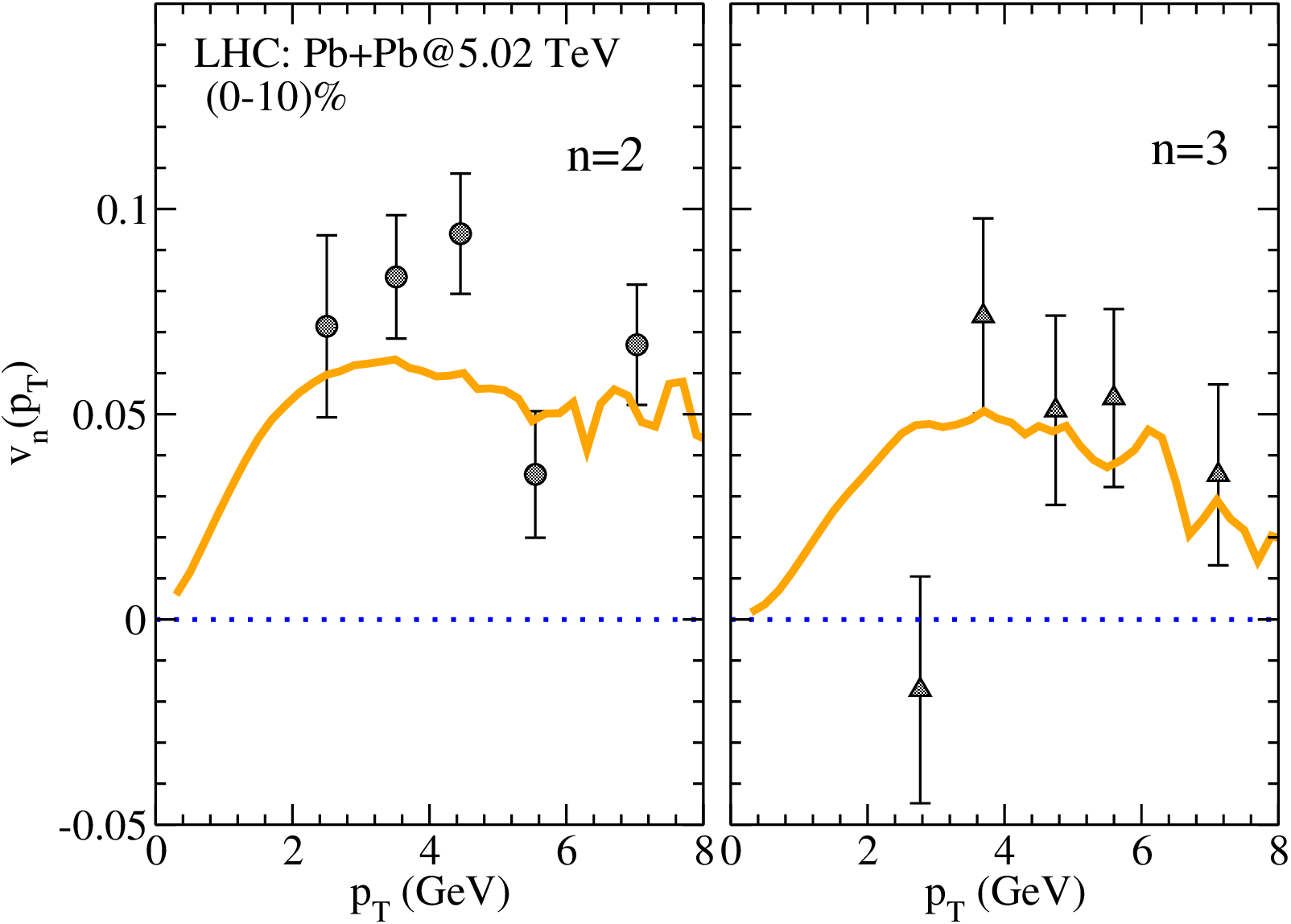}
\hspace{1.5pc}
\includegraphics[width=.35\textwidth]{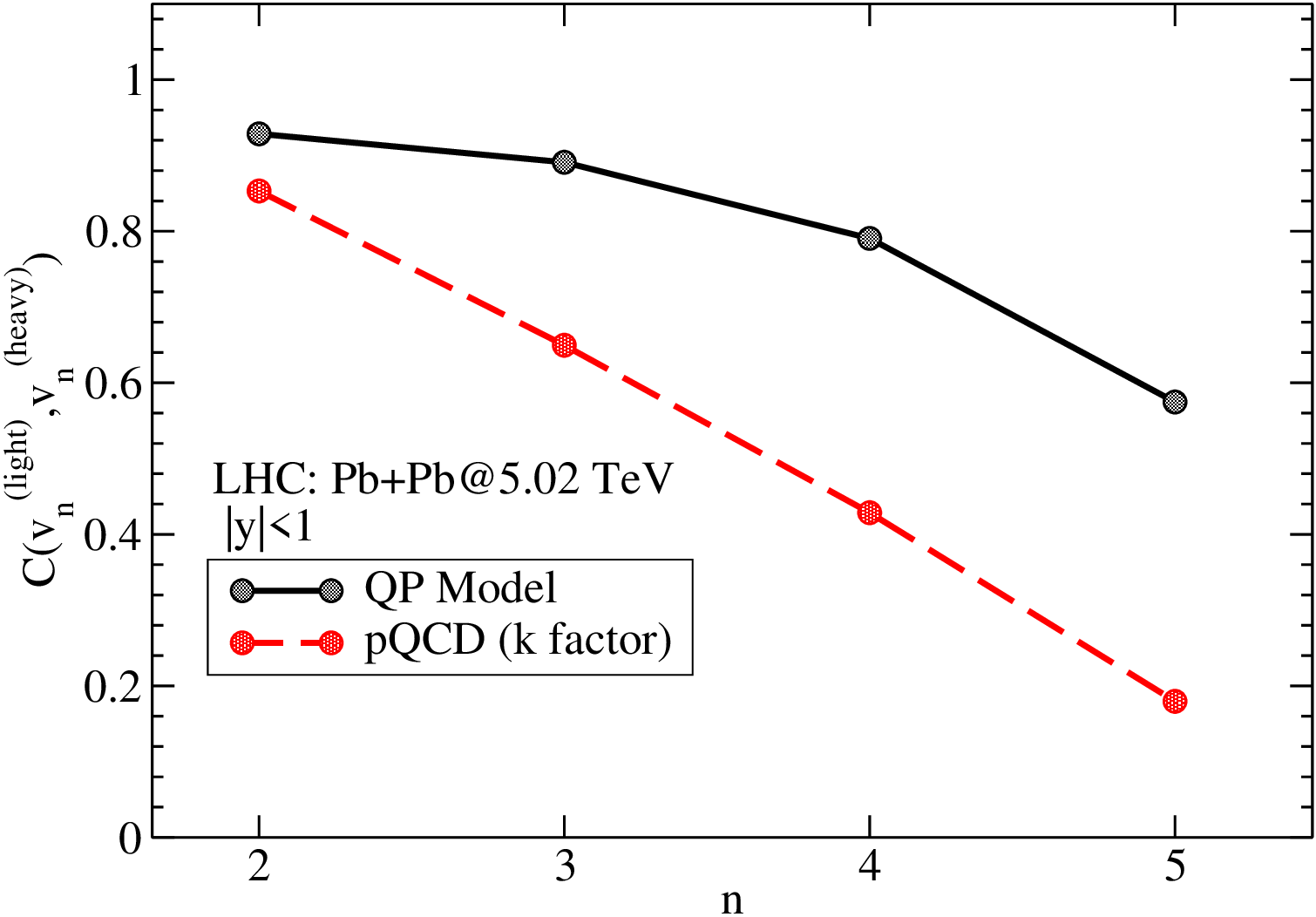}
\caption{
Left panel: $v_n(p_T)$ with $n=2,3$ for $Pb+Pb$ at $\sqrt{s_{\rm NN}}=5.02 $ TeV for $(0-10) \%$ centrality (data from\cite{Sirunyan:2017plt}).
Right panel: linear correlation coefficient $C(v_n^{light},v_m^{heavy})$ for $(0-10) \%$ centrality as a function of the order of the harmonic $n$.
Black solid line refers to QPM model while red dashed line to the case with pQCD interaction.
}\label{fig2}
\end{center}
\end{figure}
Recently, we have developed an event-by-event transport approach for the bulk in order to study the role of finite 
$\eta/s$ on the anisotropic flows $v_n(p_T)$ (see \cite{Plumari:2015cfa}).
We have used the event-by-event transport approach to extend our analysis to D meson anisotropic flow harmonics.
In the left panel of Fig.\ref{fig2} we present the D mesons $v_2(p_T)$ and $v_3(p_T)$ for central collisions in $Pb+Pb$ at 
$\sqrt{s_{\rm NN}}=5.02$ TeV. 
The anisotropic flows $v_n(p_T)$ have been calculated using the event plane method, $v_n=\langle \cos{[n(\phi -\Psi_n)]}\rangle$,
with the momentum space angles $\Psi_n=(1/n)\arctan{(\langle\sin{(n\phi)}\rangle / \langle\cos{(n\phi)}\rangle)}$.
As shown in Fig.\ref{fig2} we obtain a finite $v_3(p_T)$ which is in good agreement
with recent experimental data \cite{Sirunyan:2017plt}.  
The results shown have been obtained including only fragmentation, the inclusion of coalescence
will give an enhancement of the final anisotropic flows $v_n$.
In the right panel of Fig.\ref{fig2} it is shown the comparison of event-by-event correlations
between light flavour and heavy mesons flow harmonics at LHC energies for QPM model (black line) and pQCD interaction (red dashed line) that has a $T^2$
dependence of the drag while the QPM one has a quite weak T dependence.
The measure of the linear correlation is given by the correlation coefficient between light flavour and heavy mesons flow harmonics
\begin{equation}
C(v_n^{light},v_m^{heavy})=\frac{\sum_{i}(v_n^{i (light)}-\langle v_n^{(light)} \rangle)(v_m^{i (heavy)}-\langle v_m^{(heavy)} \rangle)}{\sqrt{\sum_{i}(v_n^{i (light)}-\langle v_n^{(light)} \rangle)^2\sum_{i}(v_m^{i (heavy)}-\langle v_m^{(heavy)} \rangle)^2}}
\end{equation}
We observe that the second and third harmonics of HQs are strongly correlated to the corresponding harmonics of light quarks with $ C(v_n^{light},v_n^{heavy}) \sim 0.9$ for $n=2,3$. 
Moreover, to highlight the impact of the interaction on $v^{light}_n-v^{heavy}_n$ correlation, we also consider pQCD interaction (red line). The $C(v_n^{light},v_m^{heavy})$ is sensitive to the heavy quarks - bulk interaction and in particular
we observe a smaller correlation for pQCD interaction, suggesting that it could give information about the T-dependence of
transport coefficients.
\section{Conclusions}
We have studied the HQ propagation in QGP at RHIC and LHC energies within a 
relativistic Boltzmann transport approach where the interaction between HQs
and light quarks of the bulk is described within 
quasi-particle model.
Within our approach we have a good description for D meson $R_{\rm AA}$ 
and $v_2$ both at RHIC and LHC energies within the experimental uncertainties.
We observe that the effect of the hadronization by coalescence is to increase the 
$R_{\rm AA}$ and $v_2(p_T)$ for $p_T > 1 $ GeV, playing a fundamental role in the simultaneous description of both $R_{\rm AA}$ and $v_2(p_T)$.
We have also studied the heavy flavour anisotropic flows within a event-by-event transport approach.
Initial state fluctuations allows to extend the analysis to high order harmonics $v_n(p_T)$
as well as to study the role of QCD interaction in developing correlations between the light and the heavy
flavour anisotropic flows. This shows how the temperature dependence of the heavy quarks - bulk interaction affect the 
heavy-light event by event $v^{light}_n-v^{heavy}_n$ correlations.






\bibliographystyle{elsarticle-num}

\end{document}